\documentclass[letterpaper]{article} % DO NOT CHANGE THIS
\usepackage[]{aaai2026}  % DO NOT CHANGE THIS
\usepackage{times}  % DO NOT CHANGE THIS
\usepackage{helvet}  % DO NOT CHANGE THIS
\usepackage{courier}  % DO NOT CHANGE THIS
\usepackage[hyphens]{url}  % DO NOT CHANGE THIS
\usepackage{graphicx} % DO NOT CHANGE THIS
\usepackage{natbib}  % DO NOT CHANGE THIS AND DO NOT ADD ANY OPTIONS TO IT
\usepackage{caption} % DO NOT CHANGE THIS AND DO NOT ADD ANY OPTIONS TO IT
\usepackage{algorithm}
\usepackage{algorithmic}
\usepackage{amssymb}
\usepackage{array}
\usepackage{booktabs}
\usepackage{graphicx}
\usepackage{tikz}
\usetikzlibrary{shapes.geometric, arrows.meta, positioning, shadows, calc, backgrounds}

\usepackage{tabularx}
\usepackage{enumitem}

\usepackage{xcolor}
\definecolor{slate}{RGB}{112, 128, 144}

\usepackage{newfloat}
\usepackage{listings}
\DeclareCaptionStyle{ruled}{labelfont=normalfont,labelsep=colon,strut=off}
\floatstyle{ruled}
\newfloat{listing}{tb}{lst}{}
\floatname{listing}{Listing}

\title{Silent Updates: Measuring and Closing the Post-Deployment Disclosure Gap}
\author{
    Sophia Abraham\textsuperscript{\rm 1},
    Ben Bucknall\textsuperscript{\rm 2}
}

\affiliations{
    \textsuperscript{\rm 1}Pivotal Research\\
    \textsuperscript{\rm 2}University of Oxford\\
    sabraha2@nd.edu%, ben.bucknall@new.ox.ac.uk
}
\begin{document}

\maketitle

\begin{abstract}
Deployed foundation models are often not static systems, with providers able to modify system behavior through fine-tuning, classifier updates, system prompt revisions, retrieval changes, and routing changes. These updates can be made \emph{silently} -- that is, without public disclosure, a version increment, or re-evaluation. Such \emph{silent updates} challenge a core assumption behind current AI governance frameworks that an externally verifiable chain of custody links the model referred to in evaluation results or a system card to the model served to users.
% Foundation model providers routinely modify deployed systems after release through fine-tunes, classifier updates, system prompt revisions, retrieval changes, and routing changes. These updates often ship \emph{silently} -- that is, without public disclosure, a version increment, or re-evaluation. Such \emph{silent updates} challenge a core assumption behind current AI governance frameworks that an externally verifiable chain of custody links the model an evaluator examines to the model users are actually served.

% In this paper, we examine post-deployment disclosure practices across nine first-party API providers and seven inference hosts. Most providers in our sample publish substantial safety documentation, including quantitative evaluations and version-specific reports. However, no provider in our sample published information allowing an external party to verify that the artifact being served is the same one that was evaluated. This points towards an ecosystem-wide lack of methods for users to reliably connect evaluation results to deployed systems.
In this paper, we examine post-deployment disclosure practices across first-party API providers and inference hosts to establish the extent to which a chain of custody exists in practice. We find that providers commonly publish substantial safety documentation, including quantitative evaluations and version-specific reports, but no provider in our sample published information allowing an external party to verify that the artifact being served is the same one referred to in this documentation.

We introduce the Silent Updates Scorecard, a public instrument for measuring post-deployment disclosure practices across providers and hosts, and preliminary results for a sample of nine first-party API providers and seven third-party inference hosts. We also propose a Three-Part Behavioral Trigger System for determining when post-deployment modifications to a system motivate disclosure or re-evaluation obligations.
% We introduce the Silent Updates Scorecard, a public instrument for measuring post-deployment disclosure practices across providers and hosts. We also present a chain-of-custody analysis of how evaluations, model identifiers, and deployed artifacts relate to one another in practice. Finally, we propose a Three-Part Behavioral Trigger System for post-deployment disclosure obligations, paired with a regulatory safe harbor for governance-focused benchmarking.
\end{abstract}

\section{Introduction}
\label{sec:introduction}
% The model evaluated in a system card or safety report is not necessarily the one deployed to users.
Foundation model deployments are not static after release and prior work has shown that deployed models continue to evolve through post-training adaptation, learning from deployment data, and ongoing operational updates \citep{ouyang2022traininglanguagemodelsfollow,xu2022learningnewskillsdeployment,obrien2023deploymentcorrectionsincidentresponse,pratt2025documenting}. In production systems, deployment-layer mechanisms such as routing policies, retrieval configuration, and system prompt revisions may further change externally observed behavior without changing the public model identifier. This motivates the need for mechanisms that link deployed behavior to the versions evaluated in safety assessments. When these changes occur without public disclosure, a version increment, or re-evaluation, we call them \emph{silent updates}. As a result, published evaluations may no longer describe the system being served, with users being unaware of this discrepancy.

This creates a problem for current AI governance frameworks. Safety evaluations, system cards, responsible scaling policies, and post-market monitoring obligations all assume that the evaluated model can be linked to the deployed one through some externally checkable chain of custody. Yet most frameworks do not clearly specify when deployment changes require disclosure or re-evaluation. Our results suggest that these links often cannot be independently verified using public documentation or API access alone.

The consequences extend beyond documentation itself. Published evaluations are increasingly used by regulators, deployers, and researchers as evidence about system behavior \citep{Mitchell_2019, EU_AI_Act_2024_Misc, NIST_AI_800-3}. But if the evaluated artifact cannot be reliably connected to the deployed system, it becomes difficult to independently verify what those evaluations apply to in practice.

We study what we term the \emph{verification gap}: the difficulty of independently connecting published evaluations to the systems users actually interact with. First, we introduce the Silent Updates Scorecard, an empirical instrument for measuring post-deployment disclosure practices across nine first-party API providers and seven inference hosts. Most providers in our sample publish extensive safety and release documentation, including quantitative evaluations and version-specific reports. At the same time, we found very limited ways for an external party to verify how those evaluations relate to the systems being served in practice. We did not identify any provider that exposes a verifiable API-to-evaluation round-trip.

Second, we examine how provider Terms of Service interact with independent evaluation. Six of nine providers in our sample include clauses that either explicitly restrict benchmarking or prohibit certain forms of competitive use in ways that may discourage external evaluation. Together, these conditions make meaningful third-party evaluation more difficult in practice \citep{longpre2024safeharboraievaluation}.

Third, we propose a Three-Part Behavioral Trigger System for determining when deployment changes should trigger disclosure obligations. The framework combines capability triggers, behavioral drift triggers, and component triggers. We also propose a regulatory safe harbor for public-interest benchmarking.

The remainder of the paper is organized as follows. 
Section 2 reviews related work on AI governance, transparency, and behavioral drift. 
Sections 3 and 4 define silent updates and describe our methodology. 
Sections 5 through 7 present the empirical results and chain-of-custody analysis. 
Sections 8 through 10 discuss governance implications, limitations, and conclusions.

\section{Background and Related Work}
\label{sec:related_work}
\subsection{Post-Deployment AI Governance}

Recent AI governance frameworks increasingly recognize the need for post-deployment monitoring and disclosure obligations for general-purpose models. The EU AI Act establishes documentation, transparency, and post-market monitoring requirements for general-purpose and systemic-risk models \citep{act2024eu}. The GPAI Code of Practice further develops these obligations through measures covering model documentation, monitoring, and reporting updates \citep{GPAICodeOfPractice2025}. These frameworks frequently rely on terms such as ``material'' or ``substantial'' change without clearly defining when disclosure or re-evaluation is required.

\subsection{Transparency Measurement}

The Foundation Model Transparency Index (FMTI) \citep{bommasani2025fmti} is the closest companion effort to ours. The 2024 and 2025 editions evaluate major foundation model developers across a broad set of transparency indicators. The 2025 report identifies post-deployment disclosure as one of the weakest areas across providers. Our scorecard focuses specifically on post-deployment change disclosure. We also evaluate inference hosts in addition to first-party providers and examine how these disclosure practices affect outside oversight.

Structured documentation has become a central mechanism for improving transparency and accountability in AI systems. Early work introduced datasheets for datasets to standardize documentation of dataset provenance, composition, intended use, and limitations \citep{gebru2021datasheets}. Subsequent work extended these ideas to model cards, system cards, and other structured disclosure formats for communicating model capabilities, limitations, and evaluation results \citep{mitchell2019modelcards, gilbert2023reward, derczynski2023assessing}. More recently, STREAM proposes standardized reporting of frontier AI capability evaluations, while Audit Cards advocate structured reporting of evaluation context and methodology to improve the interpretability and trustworthiness of AI assessments \citep{mccaslin2025stream, staufer2025audit}. Our chain-of-custody analysis extends this literature by asking whether these documentation artifacts can be reliably connected to the behavior of deployed systems over time.

\subsection{Behavioral Drift in Deployed Models}

Prior work has documented that deployed machine learning systems and commercial APIs can change substantially over time, often without explicit notification to users. Studies have observed behavioral drift in deployed language models, shifts in commercial machine learning APIs, and undocumented performance changes following model updates \citep{chen2024chatgpt, mireshghallah2024valuedrifts, gao2025modelequality, chen2021didmodelchangeefficiently, eyuboglu2024model}. More recent work has proposed efficient black-box techniques for detecting silent model updates and API changes \citep{cai2025gettingpayforauditing, chauvin2026tokenefficientchangedetectionllm, chauvin2026logprobabilitytrackingllm}. Our work is complementary: rather than developing new detection algorithms, we investigate whether existing disclosure practices allow external stakeholders to determine when changes have occurred and whether public documentation accurately reflects deployed behavior.

\subsection{Auditing and Accountability Methodology}

Our scorecard draws on the growing literature on AI auditing and accountability, particularly work positioning third-party audits as a governance mechanism for frontier AI systems \citep{raji2020closing,birhane2024ai,M_kander_2023, brundage2026frontieraiauditingrigorous}. We adopt conventions from this literature around evidence preservation, public scoring instruments, and reproducibility. The work also connects to research on AI supply chain accountability and external scrutiny ecosystems, which emphasize the need for verifiable information flows across the machine learning lifecycle and between model developers, deployers, and downstream users \citep{cobbe2023understanding,widder2023dislocated,anderljung2023publiclyaccountablefrontierllms,Ojewale_2025,spoczynski2025atlasframeworkmllifecycle}. Silent updates complicate this chain of accountability because the artifact referenced in compliance documentation may not be the one actually served downstream, weakening the connection between published evaluations and deployed systems \citep{anderljung2023publiclyaccountablefrontierllms}. 

\section{Silent Updates: Definitions and Scope}
\label{sec:taxonomy}
\subsection{Defining Silent Updates}

We define a \emph{silent update} as a change to a deployed AI system that materially affects user-facing behavior without corresponding public disclosure. The key issue is whether outside parties can observe and track those changes over time. A model update accompanied by public documentation, versioning information, or re-evaluation would not qualify as silent. By contrast, an undisclosed change to a safety classifier or routing policy would qualify as a silent update even if the underlying model weights remained unchanged.

Table~\ref{tab:defs} summarizes the operational definitions used throughout the paper. A property counts as present only when the stated condition is observable through public documentation, Terms of Service, or standard API access.

\begin{table}[t]
\centering
\small
\caption{Operational definitions of the core terms used in scoring and analysis. Each definition is evaluated only against publicly observable artifacts such as documentation, Terms of Service, and standard API responses.}
\label{tab:defs}
\begin{tabular}{p{2.5cm}p{5.0cm}}
\toprule
\textbf{Term} & \textbf{Operational definition} \\
\midrule

Snapshot binding &
A safety evaluation document names an immutable, API-callable identifier that uniquely identifies the evaluated artifact. \\[3pt]

API-to-evaluation round-trip &
An external party can map a model's API response back to the specific evaluation artifact describing the served snapshot using only public information. \\[3pt]

Policy version-binding &
A content-policy revision is linked to a specific model snapshot rather than applying to the service as a whole. \\[3pt]

Family vs.\ snapshot identifier &
A family identifier such as ``GPT-5'' denotes a product line. A snapshot identifier denotes one immutable served artifact. Only the latter satisfies snapshot-level criteria. \\[3pt]

Deterministic control vs.\ reproducible replay &
A deterministic control such as temperature 0 or a seed parameter can reduce sampling variation within a fixed snapshot. It does not guarantee that the same snapshot remains deployed over time, which is what reproducible replay would require. \\[3pt]

Silent update &
A behavioral or system change without disclosure linking the change to a specific named version and effective date. \\

\bottomrule
\end{tabular}
\end{table}

\subsection{A Taxonomy of Post-Deployment Changes}

Post-deployment changes operate at multiple layers of the deployed stack. We distinguish three categories. \emph{Model-layer changes} include weight updates from post-training, distillation, or quantization. \emph{System-layer changes} include modifications to system prompts, classifiers, retrieval components, tool access, routing policies, inference budgets, and tokenizers. \emph{Surface-layer changes} include differences between chatbot and API behavior, as well as mid-session model substitution.

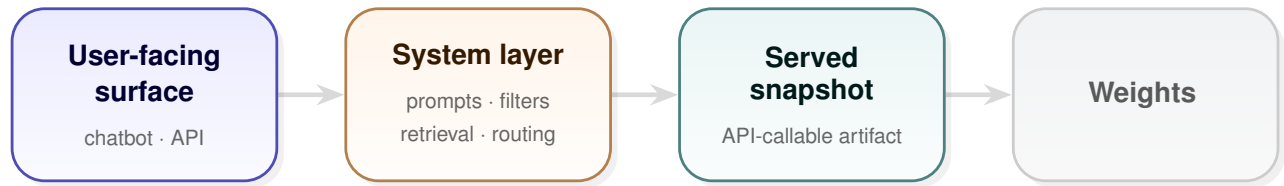
\begin{figure*}[t]
\centering
\resizebox{0.95\textwidth}{!}{%
\begin{tikzpicture}[
    font=\sffamily,
    node distance=1.0cm,
    >=Stealth,
    /utils/exec={\definecolor{local-slate}{RGB}{112, 128, 144}},
    surface-bg/.style={top color=blue!8, bottom color=blue!2},
    surface-border/.style={draw=blue!40!gray},
    surface-text/.style={text=blue!40!black},
    system-bg/.style={top color=orange!8, bottom color=orange!2},
    system-border/.style={draw=orange!40!gray},
    system-text/.style={text=orange!50!black},
    model-bg/.style={top color=teal!8, bottom color=teal!2},
    model-border/.style={draw=teal!40!gray},
    model-text/.style={text=teal!40!black},
    weights-bg/.style={top color=local-slate!10, bottom color=gray!5},
    weights-border/.style={draw=gray!40},
    basebox/.style={
        rectangle,
        rounded corners=14pt,
        minimum width=4.0cm,
        minimum height=2.6cm,
        align=center,
        line width=1.2pt,
        drop shadow={opacity=0.06, shadow xshift=3pt, shadow yshift=-3pt, fill=black}
    },
    header/.style={font=\sffamily\bfseries\large, yshift=0.5cm}
]

\node[basebox, surface-bg, surface-border] (surface) {
    {\Large\bfseries\color{blue!20!black} User-facing}\\[0.6ex]
    {\Large\bfseries\color{blue!20!black} surface}\\[1.5ex]
    {\small\color{gray!80!black} chatbot $\cdot$ API}
};
\node[header, above=of surface, surface-text] {surface-layer changes};

\node[basebox, system-bg, system-border, right=of surface] (system) {
    {\Large\bfseries\color{orange!20!black} System layer}\\[1.5ex]
    {\small\color{gray!80!black} prompts $\cdot$ filters}\\[0.6ex]
    {\small\color{gray!80!black} retrieval $\cdot$ routing}
};
\node[header, above=of system, system-text] {system-layer changes};

\node[basebox, model-bg, model-border, right=of system] (model) {
    {\Large\bfseries\color{teal!20!black} Served}\\[0.6ex]
    {\Large\bfseries\color{teal!20!black} snapshot}\\[1.5ex]
    {\small\color{gray!80!black} API-callable artifact}
};
\node[header, above=of model, model-text] {model-layer changes};

\node[basebox, weights-bg, weights-border, right=of model] (weights) {
    {\Large\bfseries\color{gray!70!black} Weights}
};

\begin{scope}[on background layer]
    \draw[->, line width=2pt, color=gray!30] (surface.east) -- (system.west);
    \draw[->, line width=2pt, color=gray!30] (system.east) -- (model.west);
    \draw[->, line width=2pt, color=gray!30] (model.east) -- (weights.west);
\end{scope}

\end{tikzpicture}%
}
\caption{The deployed AI system stack. Many post-deployment changes occur at the surface and system layers, while current governance obligations primarily attach to the model or weight layer.}
\label{fig:stack}
\end{figure*}

Current governance frameworks primarily focus on the model layer. Many deployment changes, however, occur at the system layer. Figure~\ref{fig:stack} situates these categories within the deployed stack.

\subsection{Scope of This Paper}

Our empirical analysis is limited to what can be observed externally. We use public model listings exposed through provider APIs, provider documentation and archived versions of that documentation, and behavioral fingerprints obtained through standard API calls. We do not claim visibility into internal deployment infrastructure, model registries, or proprietary system prompts and classifiers.

This limitation is central to the governance problem we study. Downstream deployers, regulators, and independent evaluators can only act on information that is externally available. This paper focuses on the gap between deployed behavior and what outside parties can independently verify.

\section{Methodology}
\label{sec:methodology}
\subsection{Provider and Inference Host Selection}

We score nine first-party API providers and seven inference hosts. The first-party providers are OpenAI, Anthropic, Google, Meta, xAI, Mistral, DeepSeek, Cohere, and AI21. The inference hosts are AWS Bedrock, Azure OpenAI, Google Vertex AI, Together AI, Groq, Fireworks AI, and Replicate.

We included first-party providers if they operate a public API serving at least one foundation model, publish at least one form of safety documentation, and plausibly fall under EU AI Act GPAI obligations based on public information about model scale. We included inference hosts if they serve at least three foundation models from distinct first-party providers and operate a public commercial API. We exclude open-weight releases without an associated hosted API because silent updates, as defined here, are a phenomenon of hosted deployment.

\begin{table}[t]
\centering
\caption{Scorecard sections and dimensions measured.}
\label{tab:sections}
\begin{tabular}{p{0.9cm}p{2cm}p{4.5cm}}
\hline
Codes & Section & What it measures \\
\hline
V1-4 & Versioning & Whether a provider exposes stable version identifiers \\
C1-5 & Changelog & Whether changes are publicly documented \\
X1-2 & Cross-surface & Whether chatbot and API model identity align \\
D1-6 & Deprecation & Whether retirement is governed by policy \\
M1-7 & Safety eval traceability & Whether published evaluations link to deployed versions \\
R1-3 & Reproducibility & Whether deterministic controls support reproducible replay \\
T1-2 & Monitoring support & Whether third-party drift detection is feasible \\
\hline
\end{tabular}
\end{table}

\subsection{The Scorecard Instrument}

The scorecard contains twenty-nine questions organized into seven sections, shown in Table~\ref{tab:sections}. These cover versioning, changelog coverage, cross-surface transparency, deprecation policy, safety evaluation traceability, reproducibility infrastructure, and monitoring support. Four additional questions apply only to inference hosts. The full question text and scoring guide appear in the supplementary materials.

The question set draws from disclosure mechanisms described in the GPAI Code of Practice, post-deployment indicators used in the Foundation Model Transparency Index \cite{bommasani2025fmti}, and Article 25 reclassification criteria. The rubric was finalized before scoring and is included in the supplementary materials in its pre-scoring form.

Each question has a literal answer and a normalized score. Binary questions score 0 or 1. Numeric questions are banded from 0 to 3 using thresholds defined in the scoring guide. List questions score by the count of items returned. Cross-surface criteria (X1--X2) were evaluated only for providers operating both a public API and a consumer-facing chatbot interface. Providers without a comparable chatbot surface were marked as not applicable (N/A) and excluded from the applicable maximum score rather than receiving zero points. The maximum first-party score is 37 points for providers where all criteria apply.

Scoring was performed by a single rater using public documentation and standard API access. Each score in the final scorecard is linked to an evidence row containing the source URL, access date, and a verbatim passage from the relevant documentation or API response.\footnote{The complete scorecard, evidence tables, and dataset are available at \url{https://github.com/sabraha2/silent-updates-scorecard}.
}

The evidence base contains more than 270 rows across system cards, model cards, technical reports, changelogs, deprecation tables, Terms of Service, and API reference pages. When current documentation was insufficient or had changed, we reconstructed historical state using Internet Archive snapshots at four points before the scoring date: twelve months, six months, three months, and one month.

% Each score in the final scorecard is linked to documentary evidence, including the source URL, access date, and verbatim supporting text.
After the initial scoring pass, an independent co-author performed an evidence audit of the completed scorecard. Rather than rescoring providers, the audit evaluated whether the cited evidence justified each assigned score, whether similar evidence was scored consistently across providers, and whether any factual inconsistencies or ambiguous applications of the rubric remained. Any flagged cases were revisited against the original documentation before the scorecard was finalized.

% This evidence record is intended to support independent verification. A reviewer or third party can confirm or contest any individual score against the source material without redoing the full scoring exercise. We treat independent replication as a priority for follow-up work. Scores for V4 and R3 were derived from documented API references rather than live API calls, so response schemas may have changed between documentation and writing.

For every question scored zero across the sample, we treated the zero as a claim to be falsified rather than asserted. We searched provider documentation, archived snapshots, API and SDK references, and system-card appendices for counterexamples satisfying the operational definitions in Table~\ref{tab:defs}. Near-matches that failed one or more criteria were recorded but not scored as satisfying the criterion. The closest near-matches are discussed in the chain-of-custody analysis.

\subsection{Chain-of-Custody Protocol}

For each first-party provider, we identified the most recent public safety evaluation document. This could be a system card, model card, technical report, or responsible scaling commitment. We then attempted to link that evaluation to the model version currently served.

We coded three dimensions: whether the evaluation named a unique identifier for the evaluated artifact, whether that identifier remained API-accessible, and whether the document provided enough methodological detail for external re-evaluation.

Each evaluation falls into one of four outcomes: chain established, version unspecified, version inaccessible, or methodology underspecified. A chain is established only when all three conditions are met.

\subsection{What We Deliberately Do Not Measure}

We restrict our empirical claims to externally observable disclosure. We do not measure weight changes directly because weight access is not available for closed-API providers. We do not measure system prompt content because production system prompts are not externally inspectable. We do not measure routing or load balancing because route assignment is not disclosed in API responses for any provider in our sample.

\section{Results}
\label{sec:results}
We present aggregate Scorecard results across the nine first-party providers and seven inference hosts in our sample. The mean first-party score is 16.4 points, corresponding to 44.4\% of the maximum applicable score under the rubric. Scores range from 8 (Meta) to 23 (OpenAI). Table~\ref{tab:rankings} presents the full first-party ranking, and Table~\ref{tab:host_rankings} summarizes inference host transparency. Table~\ref{tab:overview} summarizes the highest-level disclosure and verification properties across providers, and Table~\ref{tab:heatmap} shows the specific properties most relevant to our argument.

\begin{table}[t]
\centering
\small
\caption{First-party Scorecard totals ranked by overall score. Maximum possible score is 37; providers without applicable cross-surface criteria were normalized against their applicable maximum.}
\label{tab:rankings}
\begin{tabular}{rlcc}
\toprule
Rank & Provider & Score & \% \\
\midrule
1 & OpenAI    & 23 & 62.2 \\
2 & Cohere    & 22 & 62.9 \\
3 & Anthropic & 20 & 54.1 \\
4 & Google    & 18 & 48.6 \\
5 & xAI       & 15 & 40.5 \\
5 & Mistral   & 15 & 40.5 \\
5 & DeepSeek  & 15 & 40.5 \\
8 & AI21      & 12 & 34.3 \\
9 & Meta      &  8 & 21.6 \\
\bottomrule
\end{tabular}
\end{table}

\begin{table}[t]
\centering
\small
\caption{Inference host transparency scores. Hosts were evaluated using the hosting-specific criteria defined in the supplementary material.}
\label{tab:host_rankings}
\begin{tabular}{rlcc}
\toprule
Inference Host & Score & \% \\
\midrule
AWS Bedrock      & 3/4 & 75.0 \\
Azure OpenAI     & 3/4 & 75.0 \\
Google Vertex AI & 3/4 & 75.0 \\
Together AI      & 3/4 & 75.0 \\
Groq             & 3/4 & 75.0 \\
Replicate        & 3/4 & 75.0 \\
Fireworks AI     & 2/4 & 50.0 \\
\bottomrule
\end{tabular}
\end{table}

Inference hosts showed relatively limited variation under the hosting-specific rubric. Six of seven evaluated hosts satisfied three of four transparency criteria, with the primary limitations arising from incomplete disclosure of deployment-level transformations such as quantization and upstream model changes. Fireworks AI was the only host scoring below this threshold, reflecting gaps in lifecycle and deployment disclosure.

\begin{table}[t]
\centering
\small
\caption{Results overview. Safety evaluation artifacts are common, but the primitives needed to bind those artifacts to deployed API behavior are absent or rare.}
\label{tab:overview}
\begin{tabular}{p{4.5cm}c}
\toprule
\textbf{Observed property} & \textbf{Providers} \\
\midrule
Expose versioned or dated identifiers & 9/9 \\
Publish quantitative safety metrics & 7/9 \\
Publish per-version safety comparisons & 6/9 \\
Document snapshot-pinnable identifiers & 5/9 \\
Publish alias-to-snapshot mapping & 2/9 \\
Name the evaluated artifact with a pinned, API-callable snapshot identifier & \textbf{1/9} \\
Enable an external API-to-evaluation round-trip & \textbf{0/9} \\
Version-bind content policy & \textbf{0/9} \\
Permit third-party benchmarking under ToS & 3/9 \\
\bottomrule
\end{tabular}
\end{table}

\begin{table}[t]
\centering
\small
\caption{Disclosure versus verification across first-party providers. Filled cells ($\blacksquare$) indicate the property is present in our scoring; $\square$ indicates absence. Anthropic alone names a pinned evaluated snapshot (M1), but no provider supports the external round-trip (M2) or version-binds policy (M6).}
\label{tab:heatmap}
\setlength{\tabcolsep}{4pt}
\begin{tabular}{l ccc | ccccc}
\toprule
 & \multicolumn{3}{c|}{\textbf{Disclosure}} & \multicolumn{5}{c}{\textbf{Binding / verification}} \\
\cmidrule(lr){2-4}\cmidrule(lr){5-9}
 & V1 & M7 & V2 & V3 & M1 & M2 & M6 & T2 \\
\midrule
OpenAI    & $\blacksquare$ & $\blacksquare$ & $\blacksquare$ & $\square$ & $\square$ & $\square$ & $\square$ & $\square$ \\
Anthropic & $\blacksquare$ & $\blacksquare$ & $\blacksquare$ & $\blacksquare$ & $\blacksquare$ & $\square$ & $\square$ & $\square$ \\
Google    & $\blacksquare$ & $\blacksquare$ & $\square$ & $\square$ & $\square$ & $\square$ & $\square$ & $\square$ \\
Meta      & $\blacksquare$ & $\blacksquare$ & $\square$ & $\square$ & $\square$ & $\square$ & $\square$ & $\blacksquare$ \\
xAI       & $\blacksquare$ & $\blacksquare$ & $\blacksquare$ & $\square$ & $\square$ & $\square$ & $\square$ & $\square$ \\
Mistral   & $\blacksquare$ & $\square$ & $\blacksquare$ & $\square$ & $\square$ & $\square$ & $\square$ & $\square$ \\
DeepSeek  & $\blacksquare$ & $\square$ & $\square$ & $\square$ & $\square$ & $\square$ & $\square$ & $\blacksquare$ \\
Cohere    & $\blacksquare$ & $\blacksquare$ & $\blacksquare$ & $\blacksquare$ & $\square$ & $\square$ & $\square$ & $\square$ \\
AI21      & $\blacksquare$ & $\blacksquare$ & $\square$ & $\square$ & $\square$ & $\square$ & $\square$ & $\blacksquare$ \\
\midrule
$n/9$     & 9 & 7 & 5 & 2 & 1 & \textbf{0} & \textbf{0} & 3 \\
\bottomrule
\end{tabular}

\vspace{1mm}

{\footnotesize
\textbf{Labels:}
V1 = versioned identifiers;
M7 = quantitative safety metrics;
V2 = snapshot-pinnable identifier;
V3 = alias-to-snapshot mapping;
M1 = evaluation names an immutable API-callable identifier;
M2 = externally verifiable API-to-evaluation round-trip;
M6 = content policy bound to version;
T2 = benchmarking permitted under Terms of Service.
Anthropic's M1 is partial (see Table~\ref{tab:m-matrix}).
}

\end{table}

The most consistent findings do not appear in the total scores but in the repeated absence of specific binding and verification properties across providers. We organize these patterns into four recurring failure modes.

\subsection{Failure Mode 1: Pinned Identifiers Without Behavioral Guarantees}

All nine providers expose dated or otherwise versioned identifiers through their APIs (V1=9/9). Five of nine document a snapshot-pinnable identifier in product documentation (V2=5/9). Only two of nine publish a centralized mapping from alias identifiers such as \texttt{gpt-5} or \texttt{claude-sonnet-4-6} to the dated snapshots those aliases currently resolve to (V3=2/9).

Several providers explicitly document alias behavior in which stable API-facing identifiers resolve to changing underlying model versions over time. DeepSeek's changelog records that \texttt{deepseek-chat} and \texttt{deepseek-coder} were retained for backward compatibility while being upgraded to DeepSeek V2.5 \citep{deepseek2024notes}. Anthropic documents that aliases point to a recommended version and ``update over time'' \citep{anthropic2026aliasdocs}. Google Vertex AI similarly states that ``the auto-updated alias of a Gemini model always points to the latest stable model'' \citep{vertex2026aliasdocs}. These cases motivate our distinction between model-family aliases and immutable snapshot identifiers.

\begin{figure*}[t]
\centering

\resizebox{0.92\textwidth}{!}{
\begin{tikzpicture}[
    font=\sffamily,
    >=Stealth,
    alias-style/.style={
        draw=blue!35,
        top color=blue!8,
        bottom color=blue!2,
        rounded corners=7pt,
        minimum width=4.8cm,
        minimum height=0.85cm,
        align=center,
        font=\sffamily\bfseries,
        line width=1pt
    },
    v-box/.style={
        draw=gray!30,
        top color=gray!4,
        bottom color=white,
        rounded corners=3pt,
        minimum width=1.18cm,
        minimum height=0.55cm,
        align=center,
        font=\sffamily\scriptsize,
        line width=0.7pt
    },
    v-active/.style={
        v-box,
        draw=blue!45,
        top color=blue!12,
        bottom color=blue!3,
        font=\sffamily\scriptsize\bfseries,
        line width=1pt
    },
    arrow/.style={
        ->,
        line width=1pt,
        color=gray!45
    },
    link-arrow/.style={
        ->,
        dashed,
        line width=1pt,
        color=blue!55
    }
]

\node[alias-style] (alias) at (6.2,1.6)
{Stable API identifier: \textcolor{blue!45!black}{\texttt{deepseek-chat}}};

\node[v-box]    (v1)  at (0,0)     {V2-0517};
\node[v-box]    (v2)  at (1.55,0)  {V2-0628};
\node[v-active] (v3)  at (3.10,0)  {V2.5};
\node[v-box]    (v4)  at (4.65,0)  {V2.5-1210};
\node[v-box]    (v5)  at (6.20,0)  {V3};
\node[v-box]    (v6)  at (7.75,0)  {V3-0324};
\node[v-box]    (v7)  at (9.30,0)  {V3.1};
\node[v-box]    (v8)  at (10.85,0) {V3.1-Term.};
\node[v-box]    (v9)  at (12.40,0) {V3.2-Exp};
\node[v-box]    (v10) at (13.95,0) {V3.2};

\draw[arrow] (v1) -- (v2);
\draw[arrow] (v2) -- (v3);
\draw[arrow] (v3) -- (v4);
\draw[arrow] (v4) -- (v5);
\draw[arrow] (v5) -- (v6);
\draw[arrow] (v6) -- (v7);
\draw[arrow] (v7) -- (v8);
\draw[arrow] (v8) -- (v9);
\draw[arrow] (v9) -- (v10);

\draw[link-arrow] (alias.south) -- ++(0,-0.45) -| (v3.north);

\node[
    font=\sffamily\scriptsize\bfseries,
    text=blue!60!black
] at (4.4,1.0)
{points to};

\end{tikzpicture}
}

\caption{Alias drift on a stable API identifier. The identifier \texttt{deepseek-chat} remained stable while the underlying served artifact changed across ten documented releases over nineteen months.}
\label{fig:alias}

\end{figure*}
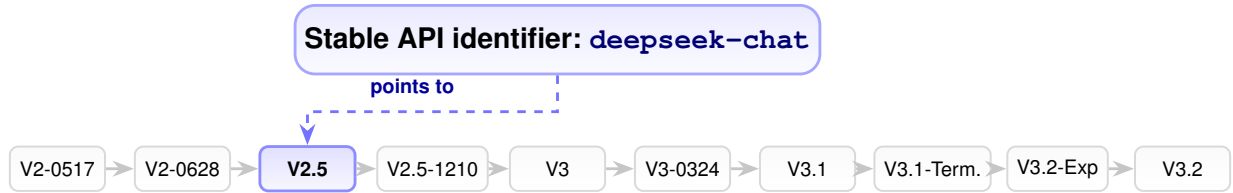

The DeepSeek case illustrates the problem clearly. The identifier \texttt{deepseek-chat} has remained continuously available since May 2024, while the underlying served model advanced through multiple versions over the same period. Figure~\ref{fig:alias} summarizes the progression recorded in DeepSeek's public release notes.

\subsection{Failure Mode 2: Changelogs Document Launches More Reliably Than Behavior Changes}

Eight of nine providers maintain public release notes or changelogs (C1=8/9). Most document new model launches and pricing changes. Far fewer document behavior changes to existing identifiers. Only four of nine publish version-level behavior change documentation (C2=3/9), and only two of nine explicitly flag safety-relevant changes in those notes (C3=2/9). Two of nine provide a machine-readable changelog feed (C4=2/9).

The asymmetry is consistent across providers. Launches and pricing changes are documented at relatively high granularity, while behavior changes to existing deployments are documented much less consistently.

\subsection{Failure Mode 3: API and Chatbot Surfaces Diverge}

Among providers operating both API and consumer chatbot surfaces, none disclose the dated snapshot identifier serving the chatbot frontend at any given time (X1=0/7). Only one provider, Mistral, propagates chatbot-side model updates into the developer-facing changelog among providers with applicable chatbot surfaces (X2=1/7).

We score X1 strictly. A family label such as ``GPT-5'' or ``Claude Sonnet 4'' does not satisfy the criterion because it does not identify the specific snapshot being served. Chatbot and API behavior may therefore differ in version, system prompt, or safety classifier configuration without external disclosure.

This creates a governance problem. A regulator observing chatbot behavior cannot reliably infer the configuration of the API service integrated downstream.

\subsection{Failure Mode 4: Safety Evaluations Are Not Bound to Served Versions}

The strongest asymmetry appears in the M-section results. Providers frequently publish safety evaluations and quantitative metrics, but none expose an externally verifiable API-to-evaluation round-trip or version-bind content policy to a specific snapshot. Anthropic partially satisfies M1 by naming a pinned evaluated snapshot, but the external verification chain still breaks at the round-trip layer.

We refer to this gap as the chain-of-custody failure and examine it in the following section.

\section{Chain-of-Custody Analysis}
\label{sec:chain_of_custody}
The M-section of the Scorecard asks whether an external party can connect three artifacts: the model identifier returned by the API, the snapshot evaluated in the safety report, and the content policy governing that snapshot. We call this connection the \emph{chain of custody}. Across the nine first-party providers in our sample, we did not identify a complete chain of custody.

The pattern holds even at providers that publish extensive safety evaluations. Figure~\ref{fig:chain} summarizes the structure of the chain and the points where it breaks across the sample.

\definecolor{warningred}{RGB}{200,30,30}

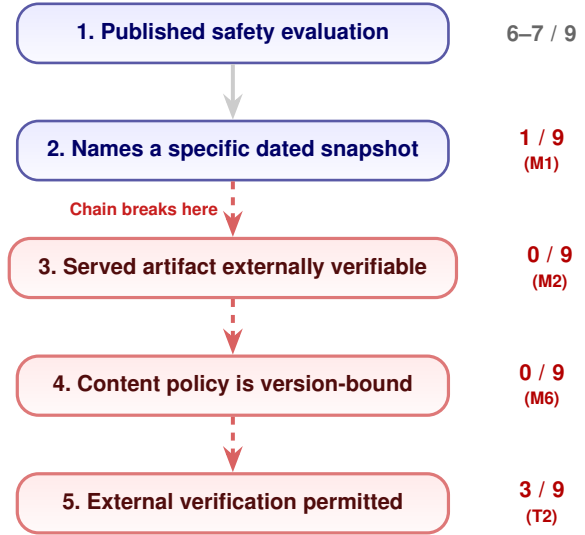
\begin{figure}[t]
\centering

\begin{tikzpicture}[
    scale=0.92,
    transform shape,
    font=\sffamily,
    node distance=0.8cm,
    >=Stealth,
    step-bg/.style={
        top color=blue!8,
        bottom color=blue!2
    },
    step-border/.style={
        draw=blue!30!gray,
        line width=1pt
    },
    break-bg/.style={
        top color=red!8,
        bottom color=red!2
    },
    break-border/.style={
        draw=warningred!60,
        line width=1.2pt
    },
    box/.style={
        rectangle,
        rounded corners=8pt,
        minimum width=6.2cm,
        minimum height=0.85cm,
        align=left,
        font=\sffamily\small\bfseries,
        inner xsep=12pt
    },
    metric/.style={
        font=\sffamily\small\bfseries,
        align=center,
        text width=1.5cm
    },
    arrow/.style={
        ->,
        line width=1.5pt,
        color=gray!40
    },
    broken-arrow/.style={
        ->,
        line width=1.5pt,
        color=warningred!70,
        dashed
    }
]

\node[box, step-bg, step-border] (eval)
{{\color{blue!40!black}1. Published safety evaluation}};

\node[box, step-bg, step-border, below=of eval] (named)
{{\color{blue!40!black}2. Names a specific dated snapshot}};

\node[box, break-bg, break-border, below=of named] (api)
{{\color{warningred!40!black}3. Served artifact externally verifiable}};

\node[box, break-bg, break-border, below=of api] (policy)
{{\color{warningred!40!black}4. Content policy is version-bound}};

\node[box, break-bg, break-border, below=of policy] (verify)
{{\color{warningred!40!black}5. External verification permitted}};

\node[metric, text=gray!80!black, right=0.45cm of eval] (m1)
{6--7 / 9};

\node[metric, text=red!70!black, right=0.45cm of named] (m2)
{1 / 9 \\ \scriptsize(M1)};

\node[metric, text=red!70!black, right=0.45cm of api] (m3)
{0 / 9 \\ \scriptsize(M2)};

\node[metric, text=red!70!black, right=0.45cm of policy] (m4)
{0 / 9 \\ \scriptsize(M6)};

\node[metric, text=red!70!black, right=0.45cm of verify] (m5)
{3 / 9 \\ \scriptsize(T2)};

\draw[arrow] (eval) -- (named);

\draw[broken-arrow] (named) -- (api)
    node[
        pos=0.5,
        left=2pt,
        font=\sffamily\scriptsize\bfseries,
        text=warningred
    ]
    {Chain breaks here};

\draw[broken-arrow] (api) -- (policy);
\draw[broken-arrow] (policy) -- (verify);

\end{tikzpicture}

\caption{The chain of custody from a published safety evaluation to an externally verifiable claim about deployed behavior. Providers commonly publish evaluation artifacts, but we did not identify the binding primitives needed to connect those artifacts to deployed snapshots.}
\label{fig:chain}

\end{figure}

\subsection{Naming Is Not Verification}

Table~\ref{tab:m-matrix} presents the per-provider M-section scores. Two binding primitives are absent across the entire sample (M2 and M6). A third, naming the evaluated snapshot (M1), is satisfied only partially by a single provider.

\textbf{M1 = 1/9.}
Most providers refer to evaluated systems by family or variant name rather than by a pinned, API-callable snapshot identifier. OpenAI's GPT-5 System Card discusses \texttt{gpt-5-thinking} and \texttt{gpt-5-main}, labels that correspond to evolving deployments over time \citep{openai2025gpt5card}. Google's Gemini 3 Pro Model Card similarly uses family-level identifiers \citep{google2025gemini3card}. Meta and xAI also rely on family designations.

Anthropic is the closest counterexample. Beginning with the Claude 4.6 generation, Anthropic documents immutable model identifiers rather than convenience aliases. These identifiers use a dateless format but correspond to fixed API-callable model snapshots\citep{anthropic2026modelids}. The Claude Opus 4.7 System Card states that evaluations were run on the final released snapshot \citep{anthropic2026opus47card}. We therefore score Anthropic as partially satisfying M1. However, some dangerous-capability metrics are reported as maxima across multiple internal snapshots, including safeguards-removed variants, so not every reported number is bound to the served artifact.

\textbf{M2 = 0/9.}
Naming a snapshot is necessary but not sufficient. We did not identify a documented mechanism at any provider that allows an external party to verify that the artifact returned by an API call is identical to the artifact described in the evaluation document. The API \texttt{model} field is provider-asserted. Even when it returns a dated identifier, no public mechanism verifies that the served weights, system prompt, and classifier configuration match the evaluated system.

\textbf{M6 = 0/9.}
Across the sample, content policy documents apply service-wide rather than to specific model versions. OpenAI's Usage Policies, Anthropic's Acceptable Use Policy, and Google's Generative AI Prohibited Use Policy all govern the provider service as a whole. We did not identify a version-scoped policy document at any provider.

\begin{table}[t]
\centering
\small
\caption{Per-provider M-section scores. Each cell is 0 (absent) or 1 (present). Anthropic partially satisfies M1 because it names a pinned evaluated snapshot but does not expose a complete external verification chain.}
\label{tab:m-matrix}

\setlength{\tabcolsep}{2.5pt}

\begin{tabular}{l|ccccccccc|c}
\toprule
 & OAI & ANT & GOO & MET & xAI & MIS & DS & COH & AI21 & $n/9$ \\
\midrule
M1 & 0 & 1 & 0 & 0 & 0 & 0 & 0 & 0 & 0 & 1 \\
M2 & 0 & 0 & 0 & 0 & 0 & 0 & 0 & 0 & 0 & \textbf{0} \\
M3 & 1 & 1 & 1 & 1 & 1 & 0 & 0 & 1 & 0 & 6 \\
M4 & 1 & 1 & 1 & 1 & 1 & 0 & 0 & 0 & 1 & 6 \\
M5 & 1 & 1 & 1 & 1 & 1 & 0 & 0 & 1 & 0 & 6 \\
M6 & 0 & 0 & 0 & 0 & 0 & 0 & 0 & 0 & 0 & \textbf{0} \\
M7 & 1 & 1 & 1 & 1 & 1 & 0 & 0 & 1 & 1 & 7 \\
\bottomrule
\end{tabular}
\end{table}

\subsection{A Worked Example: GPT-5}

GPT-5 is among the most extensively documented systems in our sample. OpenAI published a forty-page GPT-5 System Card with per-category \texttt{not\_unsafe} scores across multiple variants \citep{openai2025gpt5card}. On the M-section, it satisfies M3, M4, M5, and M7.

The document refers to the evaluated systems as \texttt{gpt-5-thinking} and \texttt{gpt-5-main}. These are family-level identifiers. OpenAI's API exposes dated snapshots such as \texttt{gpt-5-2025-08-07}, but the system card does not specify which snapshot corresponds to the evaluated deployment. An auditor querying the API therefore cannot determine which dated snapshot produced the reported safety metrics.

OpenAI's Usage Policies also apply service-wide rather than to specific snapshots, so an external party cannot determine which policy revision governed the evaluated system or whether the same policy governs the currently served deployment. OpenAI's Terms of Use additionally restrict competing-system benchmarking, limiting the external verification process that would otherwise close the loop.

The disclosures exist, but they are not externally bound to deployed artifacts.

\subsection{The Missing Primitive Is Binding}

The natural hypothesis suggested by the M2 and M6 zero counts is that providers simply do not conduct serious safety evaluations. The remaining M-section data does not support that interpretation.

Seven of nine providers publish quantitative safety metrics (M7=7/9). Six publish per-version safety comparisons across releases (M4=6/9). Six publish comparisons across simultaneously released variants (M5=6/9). Six also document substantial re-evaluation on major releases (M3=6/9).

The OpenAI GPT-5 System Card includes category-level \texttt{not\_unsafe} scores across multiple variants \citep{openai2025gpt5card}. Google's Gemini 3.1 Pro Model Card reports refusal-rate and safety deltas relative to Gemini 3 Pro \citep{google2026gemini31card}. Anthropic's Opus 4.7 System Card spans more than two hundred pages and includes comparative evaluations against earlier Opus generations. Meta's Llama 4 model card reports cyber, CBRNE, and child-safety evaluations relative to Llama 3.3 \citep{meta2025llama4card}. xAI's Grok 4.1 Model Card publishes multilingual malicious-use evaluations and explicitly corrects an error in earlier reporting \citep{xai2025grok41card}. Cohere's Command A Technical Report includes extensive distribution-based safety analysis \citep{cohere2025commanda}.

Providers are increasingly publishing substantive safety evaluations. What remains absent is a public mechanism that binds those evaluations to deployed artifacts.

The providers themselves also document the drift that makes this gap consequential. Google Vertex AI states that ``the auto-updated alias of a Gemini model always points to the latest stable model'' \citep{vertex2026aliasdocs}. Anthropic documents that convenience aliases ``update over time'' \citep{anthropic2026aliasdocs}. DeepSeek's \texttt{deepseek-chat} identifier resolved to a succession of underlying systems over nineteen months \citep{deepseek2024notes}. A persistent identifier is not a stable model.

\subsection{Cross-Surface Divergence}

The problem becomes more pronounced across hosting surfaces. We documented four recurring divergence patterns where the same nominal model identifier corresponds to materially different served behavior.

First, lifecycle divergence. Anthropic retired Claude 3 Haiku from its first-party API on 2026-04-20, while AWS Bedrock listed the corresponding deployment until 2026-09-10, a 143-day difference.

Second, content-filter divergence. Azure OpenAI applies default-on content filtering not present in OpenAI's first-party API \citep{azure2025contentfilters}, producing distinct served behavior for the same underlying \texttt{gpt-5-*} deployment.

Third, quantization divergence. Together AI is the only hosting provider in our sample that publicly discloses quantization settings at the model level (H2=1/7). The same \texttt{openai/gpt-oss-120b} weights are served at MXFP4 precision on Together AI and at undisclosed precision on Groq.

Fourth, deprecation divergence. Cohere retired \texttt{command-r-03-2024} from its first-party API on 2025-09-15, while AWS Bedrock continued serving the same model until 2026-08-19 without a public reconciliation notice.

\subsection{Contractual Limits on External Verification}

The chain-of-custody problem interacts with a second layer: contractual restrictions on independent verification.

Of the nine first-party providers in our sample, six restrict benchmarking through either explicit benchmarking clauses or broader competing-product restrictions. Cohere's Terms of Use prohibit accessing the service ``for any other benchmarking or competitive purposes'' \citep{cohere2026tos}. OpenAI, Anthropic, Google, Mistral, and xAI instead rely on broader competing-product language that can plausibly extend to benchmarking activity.

This distinction matters, but both mechanisms constrain independent verification. We therefore score T2 as absent whenever either restriction is present.

The Cohere case is particularly notable. Cohere is the highest-scoring provider in our Scorecard on transparency artifacts while simultaneously restricting external benchmarking in its Terms of Service. The possibility that competing-product clauses can be used against benchmarking efforts is not hypothetical. In July 2025, Anthropic revoked OpenAI's API access under such a clause. Contemporary reporting described benchmarking activity as among the cited concerns \citep{venturebeat2025block, wired2025block}. Together AI similarly prohibits ``competitive analysis or benchmarking'' in its Terms of Service \citep{together2026tos}.

Independent benchmarking is often proposed as a response to provider opacity \citep{bommasani2025fmti}. Yet six of nine first-party providers in our sample restrict benchmarking through either explicit benchmarking clauses or broader competing-product provisions. As a result, some of the mechanisms commonly proposed for independent verification may themselves be contractually constrained.

Providers increasingly publish detailed safety artifacts, but the technical and contractual conditions needed to independently verify those artifacts often remain limited.

\section{Behavioral Drift Evidence}
\label{sec:drift}
The chain-of-custody failure described earlier would matter less if served behavior remained stable for the lifetime of a snapshot identifier. We collect evidence to the contrary from three observable signals: provider disclosures that acknowledge deployment-layer variation, documented alias-to-snapshot transitions, and the existence of a single hosting primitive that demonstrates the feasibility of immutable snapshot binding.

\textbf{Quantization disclosure as a feasibility proof.}
Together AI is the only inference host in our sample that discloses quantization at the deployment level (H2=1/7). For models such as \texttt{openai/gpt-oss-120b}, Together specifies the served precision format directly, including MXFP4, BF16, FP8, and INT8. Groq, Fireworks AI, Replicate, AWS Bedrock, Azure OpenAI, and Google Vertex AI do not disclose equivalent information. The important observation is not the absence at six hosts but the presence at one. Snapshot-bound disclosure of a numerical precision that can affect served behavior is operationally feasible. Its absence elsewhere therefore reflects a documentation choice rather than a technical limitation.

\textbf{Documented alias-update patterns.}
Google Vertex AI states that ``the auto-updated alias of a Gemini model always points to the latest stable model'' \citep{vertex2026aliasdocs}. Groq documents a similar upgrade policy for Llama deployments, noting in release documentation that requests to legacy identifiers automatically migrate to newer variants. DeepSeek's API release history likewise shows that the stable identifier \texttt{deepseek-chat} resolved to a succession of underlying versions over time \citep{deepseek2024notes}. A persistent identifier is therefore not necessarily a persistent model. Where providers document these transitions, downstream users can at least detect that silent re-pointing may occur. Where they do not, the transition remains opaque.

\textbf{Content-hashed immutable versions.}
Replicate exposes the strongest snapshot-binding primitive observed in our sample. Each deployment is identified by a content hash, and historical hash-identified versions remain queryable through the API. An evaluator who records the hash associated with a benchmarked deployment can later verify that the served artifact is bit-identical to the evaluated one. No other provider or host in our sample exposes an equivalent primitive. The significance is broader than Replicate itself: immutable snapshot binding is not hypothetical infrastructure. It already exists in production systems.

For closed-API providers, we cannot directly tell whether behavioral changes under a stable identifier come from changes to the underlying model or deployment. What we can show is that a stable public identifier does not necessarily mean the deployment itself is stable. Quantization disclosures show that deployment-level details can be documented, alias-update policies show that stable identifiers can point to changing artifacts, and content-hashed deployments show that immutable binding is technically possible. Taken together, this strengthens the interpretation of the M-section results where the lack of a clear link between published evaluations and deployed systems does not seem to come from a fundamental technical limitation, but from the fact that these mechanisms are generally not exposed to external evaluators.

\section{Proposal: A Three-Part Behavioral Trigger System}
\label{sec:governance}
\subsection{Diagnosis}

Our results show that current disclosure practices do not provide a reliable chain of custody between published evaluations and deployed systems.

Existing governance frameworks rely on terms such as ``material change,'' ``substantial modification,'' and ``relevant changes'' to trigger disclosure or re-evaluation obligations. However, these terms are rarely tied to clear behavioral thresholds. The Commission's GPAI Guidelines include an indicative compute-based threshold for substantial modification under Article 25, but this guidance applies primarily to fine-tuning and does not address many deployment-layer modifications.

As a result, current frameworks provide limited guidance for determining when post-deployment changes should be disclosed, documented, or re-evaluated.

\subsection{Three Trigger Types}

We propose replacing ambiguous ``material change'' language with three behavioral trigger types linked to different disclosure obligations.

The numerical thresholds below are illustrative and intended only as examples of how such a system could be operationalized under existing governance frameworks.

\textbf{Capability Triggers.}
A change is treated as substantial when it produces a capability shift beyond a defined threshold on a benchmark suite, significantly changes refusal behavior on a fixed safety battery, or introduces a previously absent dangerous capability. Capability triggers initiate full re-evaluation under Article 55 and possible reconsideration of Article 25 classification status.

\textbf{Drift Triggers.}
A change is treated as documentable when behavioral fingerprints drift beyond a calibrated threshold across a large prompt battery. Examples include embedding-distribution shifts over fixed prompts or measurable changes in structured-output compliance under stable identifiers. Drift triggers initiate documentation updates under the transparency provisions of the GPAI Code of Practice.

\textbf{Component Triggers.}
A change is treated as disclosable when deployment-layer components are modified, including system prompts, classifiers, retrieval systems, routing policies, tool access, inference-compute budgets, or tokenizers. Component triggers initiate logged disclosure obligations within a fixed reporting window.

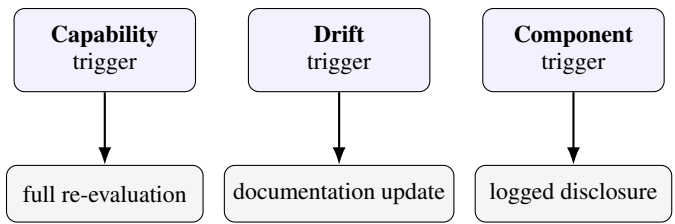
\begin{figure}[t]
\centering
\begin{tikzpicture}[
  font=\footnotesize,
  >=Latex,
  trig/.style={
      draw,
      rounded corners=4pt,
      minimum width=2.4cm,
      minimum height=1.1cm,
      align=center,
      fill=blue!5
  },
  result/.style={
      draw,
      rounded corners=4pt,
      fill=gray!8,
      minimum width=2.6cm,
      minimum height=0.75cm,
      align=center
  },
  arrow/.style={-{Latex}, thick}
]

\node[trig] (cap) at (0,0)
{\textbf{Capability}\\trigger};

\node[trig] (drift) at (3.1,0)
{\textbf{Drift}\\trigger};

\node[trig] (comp) at (6.2,0)
{\textbf{Component}\\trigger};

\node[result] (capo) at (0,-1.9)
{full re-evaluation};

\node[result] (drifto) at (3.1,-1.9)
{documentation update};

\node[result] (compo) at (6.2,-1.9)
{logged disclosure};

\draw[arrow] (cap) -- (capo);
\draw[arrow] (drift) -- (drifto);
\draw[arrow] (comp) -- (compo);

\end{tikzpicture}

\caption{
The proposed Three-Part Behavioral Trigger System.
Capability triggers initiate re-evaluation,
drift triggers initiate documentation updates,
and component triggers initiate deployment-level disclosure.
}
\label{fig:triggers}

\end{figure}

Component triggers are particularly important because many deployment changes occur outside the model weights themselves. Current governance obligations primarily attach to the model artifact, while users interact with a larger deployed system. Figure~\ref{fig:triggers} summarizes the three trigger types and their associated obligations.

\subsection{Scope and Limits}

The Three-Part Trigger System addresses several of the failures identified in our analysis. Capability triggers bind evaluations to snapshots. Drift triggers detect behavioral movement under stable identifiers. Component triggers surface deployment-layer modifications that may otherwise remain undisclosed.

The proposal does not resolve the contractual restrictions discussed earlier in the paper. A governance regime that requires disclosure while permitting providers to restrict external measurement creates tension between transparency obligations and independent verification.

We therefore pair the trigger proposal with a narrower recommendation: a regulatory safe harbor for third-party benchmarking conducted in good faith for governance, auditing, or academic research purposes. Comparable mechanisms already exist in areas such as copyright, financial regulation, and whistleblower protection. Eligibility could require methodological transparency, non-commercial intent, and provider notification within a fixed disclosure window.

\section{Limitations and Discussion}
\label{sec:limitations}
\subsection{Limitations}

Our measurement scope is bounded by what providers expose through public APIs and documentation. System prompts not surfaced through APIs, internal routing policies, and deployment-layer changes visible only through weight access remain outside our empirical reach. These constraints are themselves part of the problem we study: external governance mechanisms can rely only on information providers choose to expose publicly. 

The Scorecard was coded by a single rater. To improve consistency, the completed scorecard subsequently underwent an independent evidence audit focused on evidence-to-score consistency and factual accuracy. While this process strengthens confidence in the final scores, formal inter-rater reliability remains future work.

The proposed trigger system also depends on normative choices about which behaviors matter and how drift thresholds should be calibrated. We do not attempt to resolve those questions here.

Any threshold-based trigger system is vulnerable to Goodhart effects. Providers may optimize for reference batteries while drifting outside them. Rotating benchmark suites or granting qualified third parties access to non-public probes may mitigate this problem but will not eliminate it.

Our proposal applies most directly to hosted APIs, where providers control deployment and can therefore be held to disclosure obligations. Open-weight ecosystems complicate this logic because downstream deployers may substantially modify released models. Extending the framework to open-weight ecosystems remains future work.

While this paper focuses on the cross-sectional Scorecard and chain-of-custody analyses, future work could extend this framework with longitudinal disclosure-gap metrics that quantify how deployment evolves relative to public documentation over time. Candidate measures include coverage gaps (the proportion of deployed versions lacking corresponding changelog entries), persistence gaps (the proportion of retired versions lacking deprecation notices), and granularity gaps (the ratio of deployment changes to documented behavioral updates). Estimating these quantities reliably would require continuous monitoring of provider APIs, changelogs, and historical archives over extended periods rather than the point-in-time measurements presented here.

Finally, some providers were only partially observable during the measurement period. Meta's hosted Llama API, for example, remained preview- or waitlist-gated across portions of the study, limiting direct verification of several API-facing versioning and reproducibility criteria. Where direct inspection was not possible, scoring relied on publicly available documentation when available; otherwise, criteria were recorded as unknown rather than inferred. Scores should therefore be interpreted as measurements of observable deployment transparency rather than claims about providers' internal governance practices. 

\subsection{Political Economy}

Recent events help clarify the incentives surrounding post-deployment disclosure. In July 2025, Anthropic revoked OpenAI's Claude API access under a competing-product clause that contemporary reporting linked to benchmarking activity. Cohere's Terms of Use contain similarly explicit restrictions on benchmarking. These cases illustrate a broader pattern in our sample: providers may publish substantial transparency artifacts while simultaneously limiting outside evaluation through contractual controls.

Current governance frameworks encourage disclosure but rarely require providers to expose the technical or contractual conditions needed for outside parties to verify deployment behavior independently.

Providers can reasonably be expected to resist aspects of the trigger proposal introduced earlier in the paper. Operationalized thresholds increase compliance costs, create additional audit exposure, and reduce some of the flexibility currently afforded by silent updates. Numerical thresholds, reporting windows, and definitions of safety-relevant drift would likely remain subjects of regulatory negotiation.

At the same time, modification-trigger regimes are not unprecedented. Industries such as aviation, pharmaceuticals, and medical devices already operate under formal change-disclosure and re-evaluation requirements. We do not claim that AI systems present identical risks, only that post-deployment disclosure obligations are compatible with technologically dynamic industries.

The current regime leaves substantial discretion with providers over what counts as a disclosable change and who may independently evaluate those changes. One purpose of the Scorecard is to make those tradeoffs more visible.

\section{Conclusion}
\label{sec:conclusion}
In this paper, we show that it is not currently possible to connect evaluation results to deployed systems in externally verifiable ways. Across 9 AI providers providers we examined, the information available through public documentation and standard API access was generally insufficient to verify that the evaluated model was the one actually being served.

We describe this configuration as \emph{transparency without verifiability}. In response, we propose two complementary mechanisms: a Three-Part Behavioral Trigger System for determining when post-deployment changes should trigger disclosure obligations, and a regulatory safe harbor for governance-purposed benchmarking to support independent evaluation.

Current AI governance frameworks implicitly assume a chain of custody between evaluated and deployed systems. Our findings suggest that this chain is often missing in practice. Building more reliable links between evaluations and deployed behavior should therefore be treated as a priority in future work on AI evaluations and governance.

\section*{Ethical Statement}

This work evaluates the public disclosure practices of foundation model providers using publicly accessible documentation, Terms of Service, and standard API interactions. We do not collect personal data or rely on non-public information. Each score in the Scorecard is linked to documentary evidence to support external review and replication.

The Scorecard measures externally observable disclosure rather than underlying provider conduct. A low score reflects the absence of publicly verifiable disclosure, not evidence of misconduct or deceptive intent. Throughout the paper, we limit our claims to what can be established from public-facing artifacts and standard developer access.

The governance proposals discussed in this paper would impose additional disclosure and compliance obligations on providers if implemented. The specific thresholds and trigger mechanisms we outline are intended as illustrative starting points rather than fixed technical standards.

\bibliography{aaai2026}

\clearpage
\appendix

\section{Complete Scorecard}
\label{app:scorecard}

This appendix presents the complete scorecard rubric,
including the evaluation questions, scoring clarifications, evidence
collection methodology, and replication notes. The accompanying
spreadsheet in the supplemental materials provides provider-level scores, evidence rows, source URLs,
access dates, and scoring rationales.

\subsection{Scorecard Questions}

\begingroup
\small
\setlength{\tabcolsep}{3pt}
\renewcommand{\arraystretch}{1.08}

% ----------------------------------------------------------------------
\medskip
\noindent\textbf{Versioning (V)}
\par\nobreak\smallskip

\noindent
\begin{tabularx}{\linewidth}{
@{}
>{\centering\arraybackslash\bfseries}p{0.60cm}
>{\raggedright\arraybackslash}X
@{}
}
\toprule
ID & \textbf{Question} \\
\midrule
V1 & Do model names include version or date identifiers? \\
V2 & Is it possible to pin an exact model version that will not silently change? \\
V3 & If aliases exist (e.g., ``latest''), is it documented which exact version they reference? \\
V4 & Does the API return a system fingerprint or version identifier in responses? \\
\bottomrule
\end{tabularx}

% ----------------------------------------------------------------------
\medskip
\noindent\textbf{Changelog (C)}
\par\nobreak\smallskip

\noindent
\begin{tabularx}{\linewidth}{
@{}
>{\centering\arraybackslash\bfseries}p{0.60cm}
>{\raggedright\arraybackslash}X
@{}
}
\toprule
ID & \textbf{Question} \\
\midrule
C1 & Does a dedicated changelog page exist? \\
C2 & Does the changelog describe individual model behavior changes, rather than only new model releases? \\
C3 & Does the changelog disclose safety-relevant changes (e.g., refusal behavior, safety filters, or content policy)? \\
C4 & Is the changelog machine-readable (e.g., RSS, JSON, or a structured API)? \\
C5 & Are developers notified when changes occur? \\
\bottomrule
\end{tabularx}

% ----------------------------------------------------------------------
\medskip
\noindent\textbf{Cross-Surface Transparency (X)}
\par\nobreak\smallskip

\noindent
\begin{tabularx}{\linewidth}{
@{}
>{\centering\arraybackslash\bfseries}p{0.60cm}
>{\raggedright\arraybackslash}X
@{}
}
\toprule
ID & \textbf{Question} \\
\midrule
X1 & Does the provider disclose which API model version is deployed through the chatbot interface? \\
X2 & Does the changelog include chatbot behavior changes, or only API changes? \\
\bottomrule
\end{tabularx}

% ----------------------------------------------------------------------
\medskip
\noindent\textbf{Deprecation (D)}
\par\nobreak\smallskip

\noindent
\begin{tabularx}{\linewidth}{
@{}
>{\centering\arraybackslash\bfseries}p{0.60cm}
>{\raggedright\arraybackslash}X
@{}
}
\toprule
ID & \textbf{Question} \\
\midrule
D1 & Is there a written deprecation or lifecycle policy? \\
D2 & What is the minimum notice period for generally available model deprecation, in days? \\
D3 & What is the minimum notice period for preview or experimental model deprecation, in days? \\
D4 & Are old and new model versions available simultaneously during a transition period? \\
D5 & Are lifecycle terms formally defined (e.g., active, deprecated, or retired)? \\
D6 & How long, on average, are models supported before retirement, in days? \\
\bottomrule
\end{tabularx}

% ----------------------------------------------------------------------
\medskip
\noindent\textbf{Safety Evaluation Traceability (M)}
\par\nobreak\smallskip

\noindent
\begin{tabularx}{\linewidth}{
@{}
>{\centering\arraybackslash\bfseries}p{0.60cm}
>{\raggedright\arraybackslash}X
@{}
}
\toprule
ID & \textbf{Question} \\
\midrule
M1 & Does the model card specify the exact evaluated model version? \\
M2 & Can the deployed API model be mapped to a published safety evaluation using publicly available information? \\
M3 & When a model is updated, is the safety evaluation repeated and republished? \\
M4 & Are safety-relevant behavioral differences between model versions characterized (e.g., differences in refusal rates or blocked categories)? \\
M5 & Are safety-relevant behavioral differences between model tiers (e.g., standard versus mini) characterized? \\
M6 & Is the content or usage policy explicitly linked to specific model versions? \\
M7 & Are quantitative safety metrics published (e.g., refusal rates or category-specific scores)? \\
\bottomrule
\end{tabularx}

% ----------------------------------------------------------------------
\medskip
\noindent\textbf{Reproducibility (R)}
\par\nobreak\smallskip

\noindent
\begin{tabularx}{\linewidth}{
@{}
>{\centering\arraybackslash\bfseries}p{0.60cm}
>{\raggedright\arraybackslash}X
@{}
}
\toprule
ID & \textbf{Question} \\
\midrule
R1 & Is deterministic sampling available (i.e., does setting temperature to zero produce consistent outputs)? \\
R2 & Is a seed parameter available for reproducibility? \\
R3 & What response metadata is returned through the API (e.g., fingerprint, version, or timestamp)? \\
\bottomrule
\end{tabularx}

% ----------------------------------------------------------------------
\medskip
\noindent\textbf{Monitoring Support (T)}
\par\nobreak\smallskip

\noindent
\begin{tabularx}{\linewidth}{
@{}
>{\centering\arraybackslash\bfseries}p{0.60cm}
>{\raggedright\arraybackslash}X
@{}
}
\toprule
ID & \textbf{Question} \\
\midrule
T1 & Is a model-listing API available? \\
T2 & Do the Terms of Service permit benchmarking and publication of evaluation results? \\
\bottomrule
\end{tabularx}

% ----------------------------------------------------------------------
\medskip
\noindent\textbf{Hosting and Deployment Transparency (H)}
\par\nobreak\smallskip

\noindent
\begin{tabularx}{\linewidth}{
@{}
>{\centering\arraybackslash\bfseries}p{0.60cm}
>{\raggedright\arraybackslash}X
@{}
}
\toprule
ID & \textbf{Question} \\
\midrule
H1 & Does the hosting provider expose the underlying model version being served? \\
H2 & Does the hosting provider disclose deployment-level implementation details (e.g., quantization format)? \\
H3 & Does the hosting provider expose immutable deployment identifiers or content-hashed deployments? \\
H4 & Does the hosting provider preserve historical deployment records that are accessible through its API or documentation? \\
\bottomrule
\end{tabularx}

\endgroup

\subsection{Scoring Notes and Edge Cases}

The following rules were applied consistently across providers when
adjudicating ambiguous cases.

\begin{itemize}

\item \textbf{Evidence-collection cutoff.}
Scores reflect provider documentation, APIs, and publicly observable
artifacts available during the study's evidence-collection period.
Models, naming conventions, and documentation introduced after the
cutoff were not used to retroactively alter scores, except where updated
documentation clarified evidence that was already applicable during the
study period.

\item \textbf{V1: Version identifiers.}
Credit was awarded when publicly exposed model identifiers contained
information sufficient to distinguish meaningful model versions or
releases. Generic family aliases that could be redirected to a different
underlying model without a corresponding identifier change were not
treated as version identifiers.

\item \textbf{V2--V3: Snapshots versus aliases.}
Mutable aliases such as ``latest'' were distinguished from identifiers
that uniquely denote a particular model version or deployment. An alias
that may be silently redirected does not by itself establish a stable,
reproducible model target. Immutable snapshots, dated deployments, or
otherwise uniquely versioned identifiers satisfy this requirement when
the provider documents their stability.

\item \textbf{D2--D3: Deprecation notice periods.}
Where a provider specified different deprecation guarantees for generally
available, specialized, preview, or experimental models, the policy
applicable to the corresponding model category was used. Formal
provider-wide notice guarantees were distinguished from shorter notice
periods observed for individual specialized or experimental models.

\item \textbf{D6: Model support duration.}
Observed support duration was calculated from publicly documented model
release and retirement dates where both were available. This criterion
describes observed lifecycle duration and is distinct from a provider's
formal minimum deprecation-notice guarantee.

\item \textbf{M1: Identification of the evaluated model.}
Credit was awarded when a published safety evaluation identified the
evaluated model with sufficient specificity to distinguish it from
materially different model versions or deployments. Immutable dated
snapshots satisfy this criterion directly. Where a provider's naming
scheme does not use dated snapshots, an explicitly identified version
may satisfy the criterion if that identifier uniquely denotes the
evaluated model.

\item \textbf{M2: API-to-evaluation traceability.}
M2 required a publicly verifiable API-to-evaluation mapping. Credit was
awarded only when an external evaluator could connect the model exposed
through standard API access to a specific published safety evaluation
using publicly available information.

\item \textbf{M3: Reevaluation following model updates.}
M3 assesses whether providers establish a documented relationship
between model updates and renewed safety evaluation. Isolated examples
of reevaluation were distinguished from policies or practices indicating
that safety evaluation is systematically repeated following model
updates.

\item \textbf{M6: Version-specific policies.}
M6 required content or usage policies to be explicitly associated with
specific model versions or snapshots rather than applied only at the
service or organization level.

\item \textbf{R1: Deterministic sampling.}
The presence of a temperature parameter alone was not considered
sufficient evidence of deterministic sampling. Credit required
documentation or observable API behavior indicating that repeated
requests under deterministic settings produce consistent outputs.

\item \textbf{T2: Benchmarking permissions.}
T2 was scored as absent when the applicable Terms of Service explicitly
prohibited benchmarking or imposed restrictions that would reasonably
prevent independent third-party evaluation or publication of evaluation
results.

\item \textbf{Not-applicable criteria.}
Criteria concerning a product surface that a provider does not operate
were marked not applicable rather than scored as absent. Percentage
scores were therefore calculated against the maximum score applicable
to that provider. In particular, X1 and X2 were excluded from the
denominator for providers without a corresponding first-party chatbot
surface.

\item \textbf{Restricted API access.}
Where standard API access was unavailable or gated behind restricted
access, scoring was limited to evidence that an independent evaluator
could verify through publicly accessible documentation and available
interfaces. Restricted access was therefore treated as a limitation on
external verifiability rather than as evidence about undisclosed
provider practices.

\end{itemize}

\subsection{Evidence Collection}

Evidence was collected from publicly observable artifacts, including:

\begin{itemize}
    \item provider documentation;
    \item publicly accessible Terms of Service;
    \item API responses obtained through standard developer access;
    \item provider changelogs and release notes; and
    \item archived documentation, where applicable.
\end{itemize}

For each criterion, evidence was evaluated according to what an
independent external evaluator could verify without access to internal
provider systems. Absence of publicly observable evidence was therefore
interpreted as an absence of externally verifiable disclosure, rather
than evidence that the corresponding internal practice did not occur.

The accompanying spreadsheet provides the underlying provider-level
scores, evidence rows, source URLs, access dates, and brief scoring
rationales.

\subsection{Replication Notes}

Scoring was performed manually using publicly observable artifacts only.
The scorecard evaluates externally verifiable disclosure and
traceability practices rather than internal provider processes or
undisclosed deployment infrastructure.

The objective is to assess whether an independent evaluator can
determine which model is being served, identify meaningful changes to
that model, and verify the relationship between a deployed system and
its published safety evaluations using publicly available information.

Because provider documentation and deployed systems change over time,
the scorecard should be interpreted as a time-bounded measurement of
provider transparency rather than a permanent characterization of any
provider. The evidence-collection dates included in the accompanying
spreadsheet define the temporal scope of each assessment.

\end{document}